\documentclass[
  amsmath,amssymb,
 aps, prc, physrev,
]{revtex4-2}
\usepackage{graphicx}
\usepackage{dcolumn}
\usepackage{bm}
\usepackage{xcolor}
\usepackage{ulem}
\usepackage{cancel}

\begin{document}

\title{\textbf{Multiplicity dependence of thermal parameters in pp collisions at $\sqrt{s}=7$ TeV from statistical hadronization fits}}
\author{R.~C.~Baral}
\affiliation{Department of Physics, Ravenshaw University, Cuttack, India}
\date{\today}

\begin{abstract}
  We perform a systematic thermal analysis of identified hadron yields measured by the ALICE Collaboration in proton--proton collisions at $\sqrt{s}=7$ TeV across charged-particle multiplicity classes within the statistical hadronization model using the Thermal-FIST framework. Global fits are used to extract the chemical freeze-out temperature $T$, system volume $V$, and strangeness saturation parameter $\gamma_S$. The extracted temperature remains approximately constant at $T \simeq 155$--$165$ MeV across multiplicity, while the effective volume exhibits an approximately linear increase with event activity. In contrast, $\gamma_S$ shows a clear rise with multiplicity, indicating a progressive reduction of strangeness suppression. Derived thermodynamic quantities obtained within the model show that the energy density increases with multiplicity, while the average energy per particle increases from $\sim 0.85$ GeV to $\sim 0.99$ GeV, remaining close to $1$ GeV. Particle-to-pion ratios reproduce the strangeness-dependent hierarchy observed by ALICE.
A systematic comparison of fits constrained by hidden- and open-strangeness hadrons reveals a persistent offset in $\gamma_S$, corresponding to an approximately $4\sigma$ separation between the $\phi$- and $\Omega$-constrained results within the present quadrature-summed uncertainty treatment. These results indicate that, although high-multiplicity proton--proton collisions exhibit several features compatible with an approximate thermal description, a single global freeze-out parameterization does not fully capture the strange sector.
\end{abstract}

  \maketitle


\section{Introduction}

The statistical hadronization model (SHM) has long provided a successful and economical framework for describing hadron production in high-energy collisions~\cite{Andronic2018,Andronic2011,Cleymans1998,BraunMunzinger2004}. In this approach, the relative abundances of hadron species are determined by a thermal source characterized by a small set of parameters, most notably the chemical freeze-out temperature $T$, conserved-charge chemical potentials, and an effective volume $V$. When applied to nucleus--nucleus collisions over a wide range of energies, the SHM reproduces measured particle yields with remarkable accuracy and yields a freeze-out temperature close to the pseudo-critical temperature of quantum chromodynamics (QCD) obtained from lattice calculations~\cite{Andronic2018,Andronic2019,Becattini2001}. These observations support an interpretation in which the system formed in heavy-ion collisions approaches approximate chemical equilibrium prior to hadronization.

The extension of thermal descriptions to small collision systems such as proton--proton (pp) interactions is less straightforward. In contrast to heavy-ion collisions, pp collisions are not expected to produce large, long-lived systems, and particle production is commonly described in terms of microscopic processes such as parton fragmentation and string dynamics. The apparent success of thermal models in describing hadron yields in such small systems therefore raises important conceptual questions regarding the origin of equilibrium-like features and the role of conservation laws and phase-space constraints~\cite{Becattini2006}.

Recent measurements at the Large Hadron Collider (LHC) have renewed interest in this issue. In particular, results from the ALICE Collaboration have demonstrated that several observables traditionally associated with collective behavior in heavy-ion collisions also emerge in high-multiplicity pp and p--Pb events~\cite{ALICE2017Strangeness,ALICE2013}. A striking example is the multiplicity dependence of strange and multi-strange hadron production, where yields of strange baryons increase faster than those of non-strange hadrons with increasing event activity~\cite{ALICE2017Strangeness}. This phenomenon, commonly referred to as strangeness enhancement, exhibits a smooth evolution from pp to p--Pb and Pb--Pb systems, suggesting the presence of common underlying mechanisms.

Within the SHM framework, such behavior can be interpreted in terms of canonical suppression and its gradual reduction with increasing system size~\cite{Tounsi2003}. In small systems, exact conservation of quantum numbers, particularly strangeness, leads to a suppression of strange hadron production relative to the grand canonical limit. As the effective system size increases, this suppression weakens, and the system approaches chemical equilibrium. Deviations from full equilibration are often parametrized by the strangeness saturation factor $\gamma_S$, with $\gamma_S < 1$ indicating incomplete equilibration~\cite{Becattini2006}.

Several phenomenological studies have explored thermal model descriptions of particle production in small systems using both canonical and grand canonical approaches~\cite{Becattini2006,Andronic2018,Andronic2011,Cleymans1998}. While these studies demonstrate that thermal models can reproduce many features of the data, important open questions remain. In particular, it is not yet fully established whether a consistent set of thermal parameters can simultaneously describe hadron yields across different multiplicity classes within a single framework. Moreover, the sensitivity of thermal fits to the choice of hadron species, especially the relative role of multi-strange baryons and hidden-strangeness mesons, has not been systematically addressed. In the present work, the thermal analysis is performed within the grand-canonical ensemble with vanishing chemical potentials as a baseline phenomenological framework; it should therefore be viewed as a controlled reference description rather than a substitute for a dedicated canonical or strangeness-canonical treatment, especially in the lowest multiplicity classes.

Another aspect that deserves careful investigation is the behavior of derived thermodynamic quantities such as the energy density $\varepsilon$ and the average energy per particle $E/N$. In heavy-ion collisions, the approximate constancy of $E/N \approx 1$ GeV has been proposed as an empirical chemical freeze-out criterion~\cite{Cleymans1998,Cleymans2006}. Whether similar scaling behavior emerges in small systems, and how it correlates with the evolution of thermal parameters, provides additional insight into the degree of equilibration achieved.

In this work, we perform a comprehensive thermal analysis of identified hadron yields measured by the ALICE Collaboration in pp collisions at $\sqrt{s}=7$ TeV across charged-particle multiplicity classes~\cite{ALICE2019}. The analysis is carried out using the Thermal-FIST framework~\cite{Vovchenko2019}, which provides a modern implementation of the hadron resonance gas model including an extensive hadron spectrum and resonance decay contributions. Global fits to the measured yields are performed for each multiplicity class in order to extract the chemical freeze-out temperature $T$, the effective volume $V$, and the strangeness saturation parameter $\gamma_S$.

The primary objective of this study is to provide a systematic and internally consistent extraction of thermal parameters and derived thermodynamic quantities across multiplicity classes within a single computational framework. In addition to the primary fit parameters, we evaluate the energy density and the average energy per particle, enabling a direct comparison with empirical freeze-out criteria. We further investigate particle-to-pion yield ratios as sensitive probes of the chemical composition of the system and their dependence on multiplicity.

A key aspect of the present analysis is the examination of the sensitivity of the extracted thermal parameters to the choice of hadron species included in the fit. In particular, we compare fits constrained by multi-strange baryons, such as the $\Omega$, with those constrained by hidden-strangeness mesons, such as the $\phi$. This comparison provides insight into the extent to which a single set of thermal parameters can simultaneously describe different sectors of the strange hadron spectrum and allows us to assess possible limitations of a unified chemical freeze-out description in small systems.

The paper is organized as follows. In Sec.~II, we briefly review the statistical hadronization model and the Thermal-FIST framework. The results and their multiplicity dependence are presented and discussed in Sec.~III, where the extracted fit parameters are also summarized. Finally, conclusions are given in Sec.~IV.
\vspace{1em}

%
\section{Statistical Hadronization Model and Thermal-FIST Framework}

The statistical hadronization model (SHM) provides a successful framework for describing hadron production in high-energy collisions in terms of a hadron resonance gas in approximate thermal and chemical equilibrium~\cite{Andronic2018,Andronic2011,Cleymans1998,BraunMunzinger2004}. In this approach, the abundances of hadron species are fixed at chemical freeze-out and are characterized by a small set of thermodynamic parameters, most notably the temperature $T$, the chemical potentials associated with conserved charges, and an effective volume $V$. In the grand-canonical ensemble, the primary yield of a hadron species $i$ is given by

\begin{equation}
N_i = V \frac{g_i}{2\pi^2} m_i^2 T K_2\left(\frac{m_i}{T}\right) \gamma_S^{|S_i|},
\end{equation}
where $g_i$ is the degeneracy factor, $m_i$ is the particle mass, and $K_2$ denotes the modified Bessel function of the second kind. Here $|S_i|$ denotes the net strangeness quantum number of hadron species $i$. The parameter $\gamma_S$ accounts phenomenologically for deviations from full chemical equilibrium in the strange sector~\cite{Becattini2006}. Values of $\gamma_S<1$ indicate strangeness suppression, while $\gamma_S \rightarrow 1$ corresponds to full equilibration.

An essential ingredient of the SHM is the inclusion of resonance feeddown. Contributions from strong and electromagnetic decays of unstable hadrons substantially modify the final observed yields of stable particles and are therefore required for quantitative comparison with experiment. In the present analysis, these effects are implemented within the Thermal-FIST framework~\cite{Vovchenko2019}, which provides a modern realization of the hadron resonance gas model with an extensive particle list and the option of including resonance widths through the energy-dependent Breit--Wigner prescription~\cite{eBW}.

The choice of ensemble is particularly relevant in small collision systems. In the canonical or strangeness-canonical formulation, exact conservation of quantum numbers, especially strangeness, suppresses strange-hadron production relative to the grand-canonical limit~\cite{Tounsi2003}. A phenomenological $\gamma_S$ factor can mimic part of this behavior in grand-canonical fits, but it is not equivalent to an exact canonical treatment. This distinction is expected to be most relevant in low-multiplicity pp events, where conservation-law effects are more pronounced. In the present work, the grand-canonical ensemble with vanishing chemical potentials is used as a baseline phenomenological framework for studying multiplicity-dependent trends in the extracted parameters.

\subsection{Experimental inputs and fit implementation}

The thermal fits are performed to midrapidity integrated hadron yields measured by the ALICE Collaboration in pp collisions at $\sqrt{s}=7$ TeV~\cite{ALICE2019}. The uncertainties used in the fits correspond to the total experimental uncertainties quoted by ALICE, obtained by adding the statistical and systematic/extrapolation contributions in quadrature. For charge-conjugate hadrons, particle and antiparticle are entered as separate rows in the Thermal-FIST input table, with identical experimental values and uncertainties where applicable.

Two fit configurations are considered. In the $\phi$-constrained analysis, the fit includes $\pi^\pm$, $K^\pm$, $p$, $\bar p$, $K^0_S$, $\Lambda$, $\bar\Lambda$, $\Xi^-$, $\bar\Xi^+$, and $\phi$, while the $\Omega$ yield is excluded. In the $\Omega$-constrained analysis, the fit includes $\pi^\pm$, $K^\pm$, $p$, $\bar p$, $K^0_S$, $\Lambda$, $\bar\Lambda$, $\Xi^-$, $\bar\Xi^+$, $\Omega^-$, and $\bar\Omega^+$, while the $\phi$ meson is excluded. Since the published $\phi$ and $\Omega$ measurements are reported in multiplicity binnings that differ from those of the remaining hadrons, the non-$\phi$ and non-$\Omega$ hadron yields are recomputed, where necessary, by merging the corresponding default multiplicity classes to match the strange-hadron binning used in each fit configuration. The multiplicity classes are defined in terms of $\sigma/\sigma_{\mathrm{INEL}>0}$, where INEL$>0$ denotes the inelastic event class with at least one charged particle produced in the pseudorapidity interval $|\eta|<1$~\cite{ALICE2019}. The quoted class boundaries therefore represent the corresponding fractions of the INEL$>0$ cross section. When two or more classes are merged, the resulting yield is constructed as a weighted average,
\begin{equation}
Y_{\mathrm{merge}}=\frac{\sum_k w_k Y_k}{\sum_k w_k},
\end{equation}
where the weights $w_k$ are the widths of the corresponding $\sigma/\sigma_{\mathrm{INEL}>0}$ intervals. The corresponding point-to-point uncertainty is propagated in the same way. In the $\Omega$-constrained analysis, the corresponding multiplicity classes are I+II, III+IV, V+VI, VII+VIII, and IX+X.  For the $\phi$-constrained analysis the non-$\phi$ hadron yields are reorganized to match the nine-class $\phi$ binning. The multiplicity classes used in the two fit configurations are summarized in Table~\ref{tab:mult_classes}.
\begin{table*}[th]
\centering
\caption{
Multiplicity classes used in the thermal fits. The $\phi$-constrained analysis employs the ALICE $\phi$ multiplicity definition with nine classes, while the $\Omega$-constrained analysis uses the five published $\Omega$ multiplicity classes. In both cases, hadron yields not published in the corresponding strange-hadron binning are recomputed by merging the relevant default multiplicity classes using the widths of the $\sigma/\sigma_{\mathrm{INEL}>0}$ intervals as weights.
}
\renewcommand{\arraystretch}{1.15}
\begin{tabular}{ c c c}
\hline
 Class label & Multiplicity interval (\%) & $\langle dN_{\mathrm{ch}}/d\eta \rangle$ \\
\hline
& $\phi$-constrained & \\
\hline
  I      & 0--0.95      & 21.3 $\pm$ 0.6 \\
  II     & 0.95--4.7    & 16.5 $\pm$ 0.5 \\
  III    & 4.7--9.5     & 13.5 $\pm$ 0.4 \\
  IV+V   & 9.5--19      & 10.8 $\pm$ 0.3 \\
 VI     & 19--28       & 8.45 $\pm$ 0.25 \\
 VII    & 28--38       & 6.72 $\pm$ 0.21 \\
VIII   & 38--48       & 5.40 $\pm$ 0.17 \\
 IX     & 48--68       & 3.90 $\pm$ 0.14 \\
 X      & 68--100      & 2.26 $\pm$ 0.12 \\
\hline
& $\Omega$-constrained & \\
\hline
I+II     & 0--4.7 & 17.47 $\pm$ 0.40 \\
 III+IV   & 4.7--14 & 12.53 $\pm$ 0.35 \\
 V+VI     & 14--28 & 9.04 $\pm$ 0.27 \\
 VII+VIII & 28--48 & 6.06 $\pm$ 0.19 \\
 IX+X     & 48--100 & 2.89 $\pm$ 0.13 \\
\hline
\end{tabular}
\label{tab:mult_classes}
\end{table*}
For the $\Omega$--$\phi$ tension study, the $\phi$-side multiplicity classes are additionally combined into five matched effective points in order to reproduce as closely as possible the mean charged-particle multiplicities of the five $\Omega$ classes. The resulting $\phi$ reference values of $\langle dN_{\mathrm{ch}}/d\eta \rangle$ are $17.47 \pm 0.40$ for I+II, $12.68 \pm 0.22$ for II+III+(IV+V), $9.32 \pm 0.27$ for III+(IV+V)+VI+VII, $6.06 \pm 0.19$ for VII+VIII, and $2.89 \pm 0.13$ for IX+X.

The Thermal-FIST calculations are performed in the ideal hadron resonance gas model using the grand-canonical ensemble, quantum statistics for all hadrons, and energy-dependent Breit--Wigner resonance widths. Resonance decays are included, while weak decays are excluded in order to match the experimental treatment of the data. The fitted parameters are the temperature $T$, the radius parameter $R$, and the strangeness saturation factor $\gamma_S$. The effective volume quoted below is obtained from the fitted radius through $V=\frac{4\pi}{3}R^3$.

For each multiplicity class, the fit parameters are determined by minimizing
\begin{equation}
\chi^2 = \sum_i \frac{\left(N_i^{\mathrm{exp}}-N_i^{\mathrm{model}}\right)^2}{\sigma_i^2},
\end{equation}
where $N_i^{\mathrm{exp}}$ and $N_i^{\mathrm{model}}$ denote the experimental and model-predicted yields of species $i$, respectively, and $\sigma_i$ is the corresponding total experimental uncertainty. In the present baseline analysis, the minimization uses the total point-to-point uncertainties quoted by ALICE, obtained by adding the statistical and systematic/extrapolation contributions in quadrature. A covariance matrix for multiplicity-correlated systematic uncertainties is not included. This approximation should therefore be kept in mind when interpreting quantitative differences between the $\phi$- and $\Omega$-constrained fits.

The extracted thermal parameters are subsequently used within Thermal-FIST to evaluate derived thermodynamic quantities, including the energy density $\varepsilon$ and the average energy per particle $E/N$. The use of a single, internally consistent framework across multiplicity classes makes it possible to study the evolution of the fitted parameters and derived quantities with event activity, and to assess their sensitivity to the hadron species included in the fit.

\section{Results and Discussion}

In this section, we present the results of the thermal analysis of identified hadron yields in pp collisions at $\sqrt{s}=7$ TeV. The extracted thermal parameters are studied as functions of the charged-particle multiplicity $\langle dN_{\mathrm{ch}}/d\eta \rangle$, which serves as a proxy for the event activity and the effective system size. The analysis aims to quantify the extent to which a statistical hadronization description can account for the observed particle yields across multiplicity classes and to identify systematic trends in the extracted thermodynamic parameters.

\subsection{Fit Quality}

The quality of the thermal fits is evaluated using the reduced chi-square, $\chi^2/\mathrm{dof}$, obtained for each multiplicity class. The corresponding values are listed in Tables~\ref{tab:phi} and~\ref{tab:omega}. Across the full multiplicity range, the statistical hadronization model provides a reasonable baseline description of the measured hadron yields, with $\chi^2/\mathrm{dof}$ values typically in the range of $\sim 2$--$6$. Although the fits do not achieve perfect statistical agreement, the observed level of agreement is broadly consistent with previous thermal model analyses of hadron production in small systems, where residual deviations are expected from finite-size effects, canonical suppression, and possible non-equilibrium dynamics in the strange sector~\cite{Andronic2018,Becattini2006,Tounsi2003,ALICE2017Strangeness}. In particular, the larger $\chi^2/\mathrm{dof}$ values in some multiplicity classes may reflect tensions in simultaneously describing hadrons with different strangeness content within a single parameter set.

\begin{figure}[htb]
\centering
\includegraphics[width=0.5\linewidth]{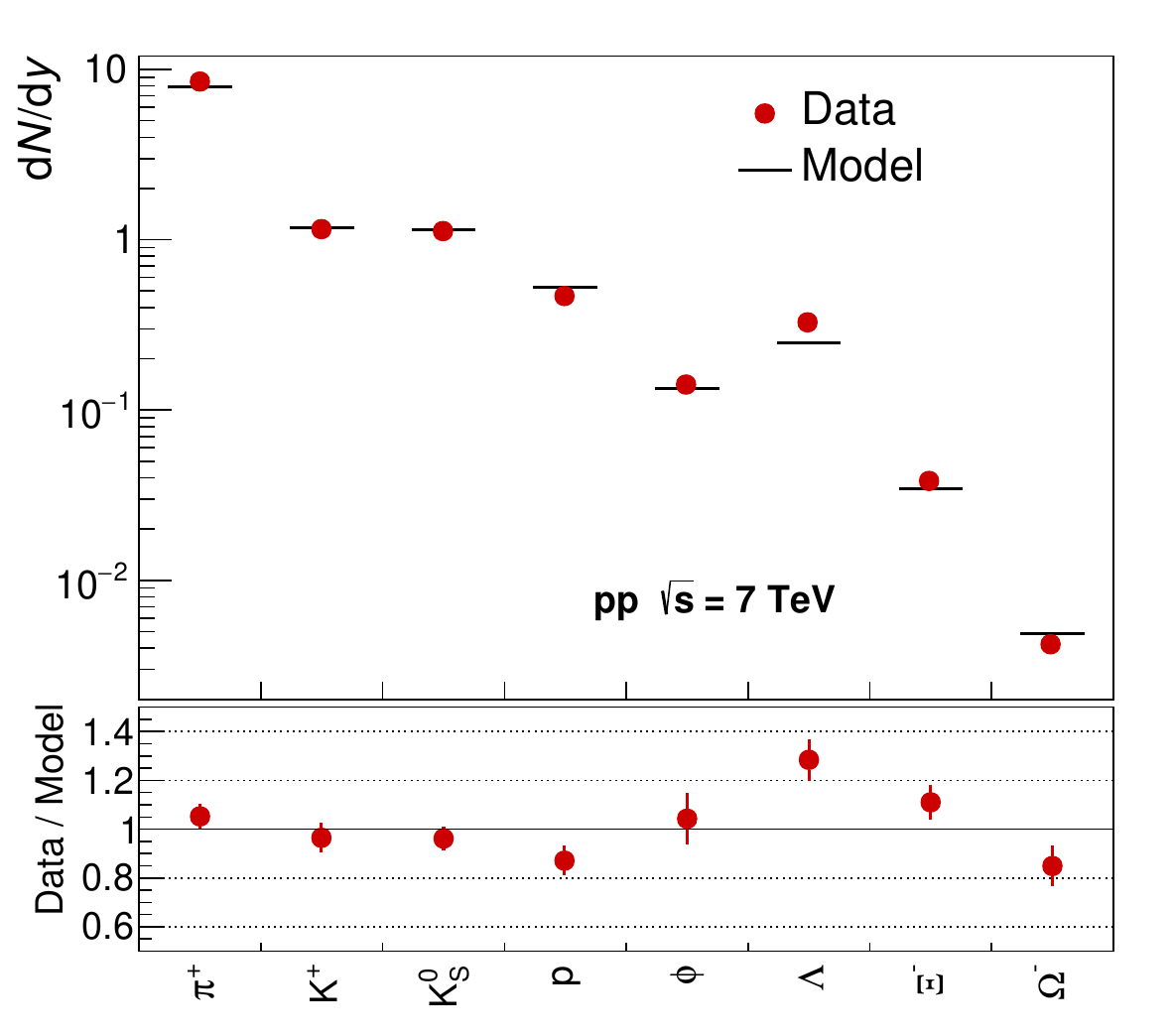}
\caption{Measured hadron yields and corresponding thermal model predictions for the combined multiplicity class I+II in pp collisions at $\sqrt{s}=7$ TeV. The upper panel shows the experimental data from ALICE and the results of global Thermal-FIST fit. The lower panel shows data-to-model ratios, with the horizontal line indicating unity. Error bars represent the experimental uncertainties.}
\label{fig:ratio}
\end{figure}

As a representative example, Fig.~\ref{fig:ratio} shows the comparison between measured hadron yields and thermal-model predictions for the combined multiplicity class I+II. In this interval, the available data include a broad set of hadron species, including multi-strange baryons and the $\phi$ meson, making it particularly useful for assessing the overall fit quality. For most species, the data-to-model ratios remain close to unity within the quoted uncertainties, indicating that the model captures the dominant pattern of hadron production. At the same time, visible deviations remain for some strange hadrons, indicating that a single global freeze-out description does not reproduce all sectors with equal accuracy. This point will become more evident in the discussion of $\gamma_S$, particle-to-pion ratios, and the $\Omega$--$\phi$ comparison below.

\subsection{Chemical Freeze-out Temperature}

\begin{figure}[htbp]
\centering
\includegraphics[width=0.4\linewidth]{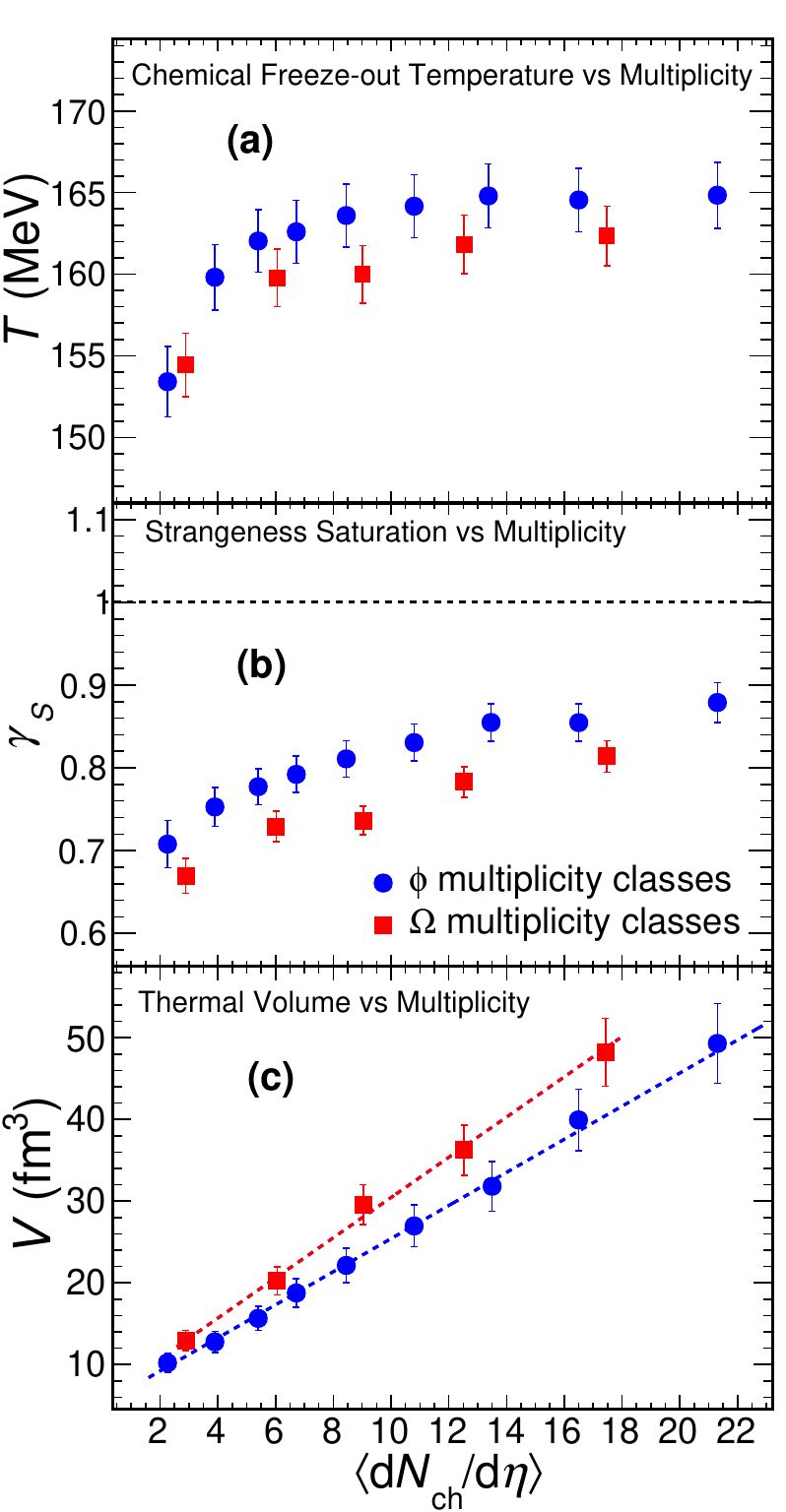}
\caption[Extracted thermal parameters as functions of multiplicity.]{Extracted thermal parameters as functions of the charged-particle multiplicity $\langle dN_{\mathrm{ch}}/d\eta \rangle$ in pp collisions at $\sqrt{s}=7$ TeV. Panels show
(a) the chemical freeze-out temperature $T$, 
(b) the strangeness saturation parameter $\gamma_S$, and 
(c) the system volume $V$. 
Results are obtained from global fits to identified hadron yields using the statistical hadronization model within the Thermal-FIST framework. Different markers correspond to fits performed using the $\phi$- and $\Omega$-constrained multiplicity classes. Error bars represent fit uncertainties. }
\label{fig:parameters}
\end{figure}

The extracted chemical freeze-out temperature $T$ as a function of charged-particle multiplicity $\langle dN_{\mathrm{ch}}/d\eta \rangle$ is shown in Fig.~\ref{fig:parameters}(a). Within uncertainties, $T$ exhibits only a weak dependence on multiplicity and remains approximately constant over the full event-activity range, with values in the interval $T \simeq 155$--$165$ MeV. A mild decrease is observed toward the lowest multiplicity classes.

The approximate constancy of $T$ suggests that the chemical freeze-out temperature is only weakly sensitive to the system size in the present data set. The extracted values are numerically close to the QCD crossover range from lattice calculations, $T_c \sim 155$--$160$ MeV~\cite{Bazavov2012,Borsanyi2010}, and are broadly consistent with previous thermal-model analyses of hadron production in both heavy-ion and small collision systems~\cite{Andronic2018,Becattini2006}. However, in the present pp analysis this agreement should be interpreted as qualitative rather than as evidence for exact thermal equilibration. The slight reduction of $T$ in the lowest multiplicity classes may reflect increasing finite-size and non-equilibrium effects in small systems. Overall, the temperature remains comparatively stable with multiplicity, in contrast to the stronger multiplicity dependence observed for $\gamma_S$ and $V$.

\subsection{Strangeness Saturation}

The strangeness saturation parameter $\gamma_S$ as a function of charged-particle multiplicity $\langle dN_{\mathrm{ch}}/d\eta \rangle$ is shown in Fig.~\ref{fig:parameters}(b). In clear contrast to the weak multiplicity dependence of $T$, $\gamma_S$ exhibits a pronounced and systematic rise with event activity. The extracted values increase from approximately $\gamma_S \simeq 0.7$ in the lowest multiplicity classes to values close to unity in the highest multiplicity events. Within the statistical hadronization framework, this trend reflects a progressive reduction of strangeness suppression with increasing system size. In small systems, exact conservation of quantum numbers, particularly strangeness, suppresses strange-hadron production relative to the grand-canonical limit~\cite{Tounsi2003}. As the charged-particle multiplicity increases, this suppression becomes weaker, and the extracted $\gamma_S$ correspondingly approaches unity.

The observed increase of $\gamma_S$ is qualitatively consistent with the strangeness-enhancement pattern measured by the ALICE Collaboration in pp and p--Pb collisions~\cite{ALICE2017Strangeness}. In particular, the increasing yields of strange and multi-strange hadrons relative to pions can be interpreted in terms of an increasing strange-quark phase-space occupancy. At the highest multiplicities, the fitted values of $\gamma_S$ indicate that the strange sector approaches approximate chemical saturation, although the residual deviations in fit quality suggest that a fully equilibrated description is not achieved uniformly for all hadron species.

A comparison of the two fit configurations further shows that the extracted $\gamma_S$ depends on the strange hadron used to constrain the fit. Fits including the open-strangeness baryon $\Omega$ systematically yield lower values of $\gamma_S$ than fits including the hidden-strangeness meson $\phi$, indicating a nontrivial species dependence in the strange sector. This behavior points to a limitation of a single global freeze-out description and anticipates the more detailed discussion of the $\Omega$--$\phi$ tension below. At the same time, the interpretation of the lowest-multiplicity values should be made with caution, since exact conservation effects are expected to be more important there than in the grand-canonical approximation employed in the present analysis.

\subsection{Volume Scaling}

The effective system volume $V$ extracted from the thermal fits as a function of charged-particle multiplicity $\langle dN_{\mathrm{ch}}/d\eta \rangle$ is shown in Fig.~\ref{fig:parameters}(c). A clear and monotonic increase of $V$ with multiplicity is observed across the full range of event activity. To quantify this dependence, the volume is fitted with a linear function of the form $V = p_0 + p_1 \langle dN_{\mathrm{ch}}/d\eta \rangle$. The resulting fits, shown as lines in Fig.~\ref{fig:parameters}(c), describe the data well within the quoted uncertainties.
For the $\phi$ multiplicity classes, the fit yields $p_0 = 5.15 \pm 1.08$ and $p_1 = 2.03 \pm 0.15$, with $\chi^2/\mathrm{dof} = 0.52/7$. For the $\Omega$ multiplicity classes, the corresponding values are $p_0 = 5.82 \pm 1.59$ and $p_1 = 2.46 \pm 0.22$, with $\chi^2/\mathrm{dof} = 0.51/3$.
 The extracted slopes are compatible within uncertainties at the $\sim 2\sigma$ level, suggesting that the overall scaling behavior is similar, although a mild sensitivity to the choice of hadron species is present. The non-zero intercept reflects the effective nature of the volume parameter within the statistical model and should not be interpreted as a direct geometric size of the system.

The approximately linear dependence of $V$ on $\langle dN_{\mathrm{ch}}/d\eta \rangle$ indicates that the effective particle-emitting volume increases with event activity. This behavior supports the interpretation of charged-particle multiplicity as a proxy for the system size and this trend is in line with the expectation that larger event activity corresponds to a larger particle-emitting source in the thermal fit~\cite{Andronic2018}.

\subsection{Energy Density and Average Energy per Particle}

\begin{figure}[htbp]
\centering
\includegraphics[width=0.65\linewidth]{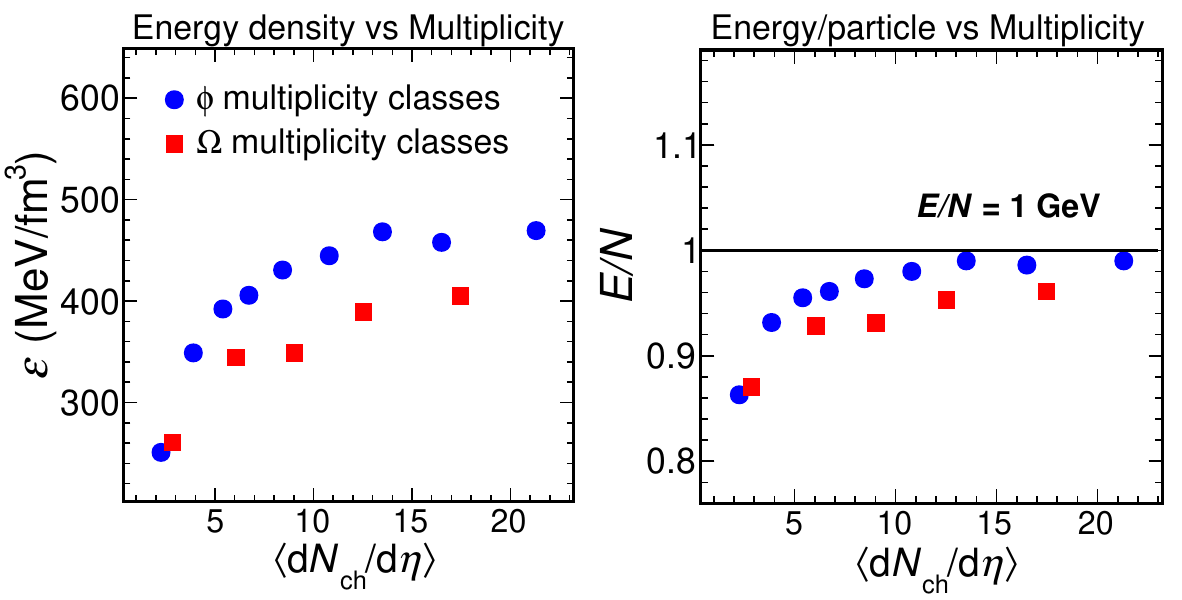}
\caption{Energy density $\varepsilon$ (left) and average energy per particle $E/N$ (right) as functions of charged-particle multiplicity $\langle dN_{\mathrm{ch}}/d\eta \rangle$ in pp collisions at $\sqrt{s}=7$~TeV. Different markers correspond to the $\phi$- and $\Omega$-constrained multiplicity classes. The solid line in the right panel indicates the empirical freeze-out condition $E/N = 1$~GeV.}
\label{fig:energy}
\end{figure}

The energy density $\varepsilon$ and the average energy per particle $E/N$ are obtained within the Thermal-FIST framework from the extracted thermal parameters as functions of charged-particle multiplicity $\langle dN_{\mathrm{ch}}/d\eta \rangle$. The results are shown in Fig.~\ref{fig:energy}. The energy density exhibits a moderate increase with multiplicity, indicating that higher-multiplicity events correspond to larger energy densities at chemical freeze-out. This behavior follows from the combined evolution of the fitted thermal parameters and the changing hadronic composition with event activity. 

The ratio $E/N$ shows a systematic increase with multiplicity, rising from $\sim 0.86$~GeV at the lowest multiplicity to $\sim 0.99$~GeV in the highest multiplicity class, while remaining close to $1$~GeV overall. This behavior is notable because it resembles the empirical freeze-out criterion observed in heavy-ion collisions, where the average energy per particle at chemical freeze-out is found to be approximately constant over a wide range of collision energies~\cite{Cleymans1998,Cleymans2006}. In the present case, however, the comparison should be regarded as qualitative, since the thermodynamic assumptions underlying the criterion are not guaranteed to hold in small pp systems. The deviations at low multiplicity may reflect the increasing importance of non-equilibrium effects and finite-size corrections.

Overall, the observed behavior of $\varepsilon$ and the near-constant value of $E/N$ are compatible with an approximate thermal description of hadron production in high-multiplicity pp collisions. At the same time, the deviations observed at low multiplicity indicate that a simple equilibrium interpretation is less reliable in the smallest systems.

\subsection{Multiplicity dependence of particle-to-pion ratios}
\begin{figure}[!hbt]
\centering
\includegraphics[width=0.35\textwidth]{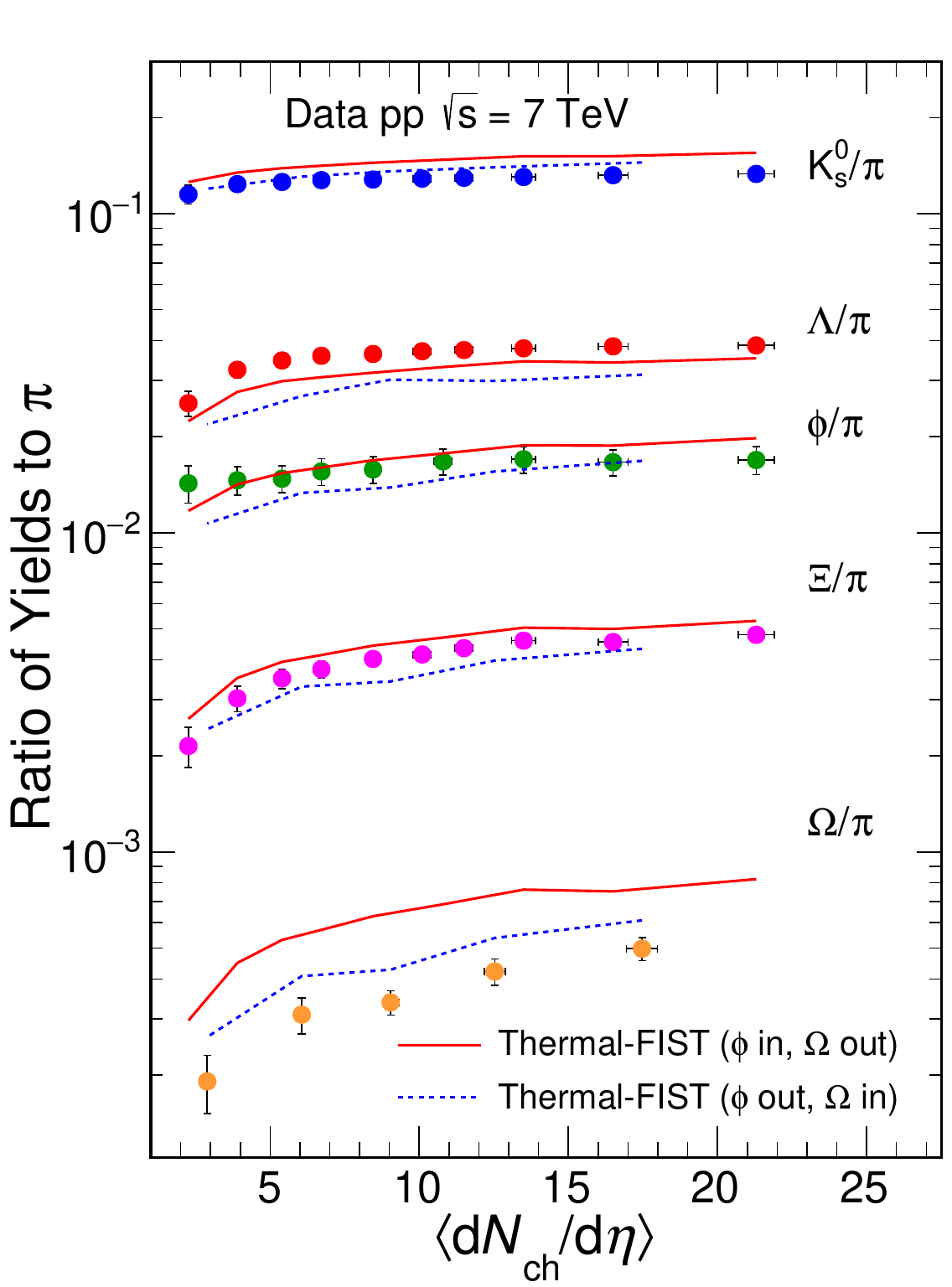}
\caption{
Multiplicity dependence of particle-to-pion yield ratios in pp collisions at $\sqrt{s}=7$ TeV. The ratios $K_S^0/\pi$, $\Lambda/\pi$, $\Xi/\pi$, $\Omega/\pi$, and $\phi/\pi$ are shown as functions of $\langle dN_{ch}/d\eta \rangle$. The experimental data (solid markers) with uncorrelated systematic uncertainties (vertical bars) are compared with thermal model calculations obtained using two different fit configurations within the Thermal-FIST framework: one including the $\phi$ meson and excluding the $\Omega$ baryon, and the other including the $\Omega$ baryon and excluding the $\phi$ meson. }
\label{fig:ratios}
\end{figure}
To further investigate the evolution of the chemical composition of the system with event activity, particle-to-pion yield ratios are studied as functions of charged-particle multiplicity $\langle dN_{\mathrm{ch}}/d\eta \rangle$. Ratios such as $K^0_S/\pi$, $\Lambda/\pi$, $\Xi/\pi$, $\Omega/\pi$, and $\phi/\pi$ provide sensitive probes of strangeness production while largely removing the trivial volume dependence of particle yields. The results are shown in Fig.~\ref{fig:ratios}.
The measured ratios exhibit a clear multiplicity ordering. The $K^0_S/\pi$ ratio varies only weakly with multiplicity, whereas the rise becomes progressively stronger for $\Lambda/\pi$, $\Xi/\pi$, and especially $\Omega/\pi$. This hierarchy, in which the multiplicity dependence strengthens with increasing strangeness content, directly reflects the strangeness-enhancement pattern measured by the ALICE Collaboration in small systems~\cite{ALICE2017Strangeness}. Within the statistical hadronization framework, such behavior is naturally associated with the progressive reduction of strangeness suppression as the event activity increases.

The $\phi/\pi$ ratio shows a more moderate increase with multiplicity compared to open-strangeness hadrons. Since the $\phi$ meson carries hidden strangeness ($s\bar{s}$) and zero net strangeness, its production is less sensitive to the strangeness saturation parameter $\gamma_S$. As a result, the $\phi$ meson provides a complementary probe of the strange sector and helps disentangle different mechanisms contributing to strangeness production.

The thermal-model calculations reproduce the overall multiplicity evolution of the ratios, but systematic differences are observed between the two fit configurations. In particular, the configuration constrained by $\Omega$ generally yields lower strange-to-pion ratios than the configuration constrained by $\phi$, reflecting the stronger sensitivity of multi-strange baryons to the strangeness occupancy parameter. The largest discrepancies appear in the multi-strange sector, indicating that all strange-hadron ratios are not reproduced with the same accuracy within a single global parameter set.

Taken together, the particle-to-pion ratios support the interpretation that strangeness suppression is progressively reduced with increasing multiplicity. At the same time, the residual tension between the hidden-strangeness and open-strangeness sectors anticipates the more detailed comparison presented in the following subsection.

\subsection{$\Omega$-$\phi$ Tension}

To quantify the sensitivity of the extracted thermal parameters to the choice of strange hadron used to constrain the fit, we compare two fit configurations: one including the hidden-strangeness meson $\phi$ while excluding the open-strangeness baryon $\Omega$, and the other including $\Omega$ while excluding $\phi$.
Since the $\phi$ and $\Omega$ measurements are reported in different multiplicity binnings, the direct $\Omega$--$\phi$ comparison is performed using five matched effective $\phi$ reference points constructed to reproduce as closely as possible the mean charged-particle multiplicities of the five $\Omega$ classes, as described in Sec.~II. For brevity, the rows in Table~\ref{tab:omega_phi_tension} are labeled by the corresponding $\Omega$ reference classes, while the $\phi$-side entries are evaluated at the matched effective $\phi$ points defined in Sec.~II.

\renewcommand{\arraystretch}{1.1}
\begin{table}[thb]
\centering
\caption{
Comparison of thermal fit parameters obtained from $\phi$-constrained (including $\phi$, excluding $\Omega$) and $\Omega$-constrained (including $\Omega$, excluding $\phi$) fits across multiplicity classes. Differences are defined as $(\phi\text{-in}) - (\Omega\text{-in})$. The quoted averages are arithmetic means over the five matched multiplicity classes; the uncertainties denote the class-to-class standard deviation. The table illustrates systematic offsets in $\gamma_S$, temperature $T$, and fit quality $\chi^2/\mathrm{dof}$ between the two configurations.
}
\begin{tabular}{c c c c c c c}
  \hline  
  Multiplicity & $\gamma_S^{\mathrm{(\phi-in)}}$ & $\gamma_S^{\mathrm{(\Omega-in)}}$  & $\Delta \gamma_S$ & $\Delta T$ (MeV) & $\Delta \chi^2/\mathrm{dof}$ \\
  
\hline
I+II     & 0.860  & 0.814  & 0.046  & 2.28  & -1.08 \\
III+IV   & 0.847  & 0.783  & 0.064  & 2.92  & -1.84 \\
V+VI     & 0.819  & 0.736  & 0.083  & 3.77  & -2.73 \\
VII+VIII & 0.784  & 0.729  & 0.055  & 2.52  & -1.46 \\
IX+X     & 0.735  & 0.670  & 0.065  & 2.51  & -0.79 \\
\hline
Average & --- & --- & $0.063 \pm 0.015$ & $2.80 \pm 0.43$ & $-1.58 \pm 0.45$ \\
\hline

\end{tabular}
\label{tab:omega_phi_tension}
\end{table}

 As shown in Table~\ref{tab:omega_phi_tension}, the comparison reveals a coherent and systematic separation between the two fit configurations across the full multiplicity range. In every matched multiplicity class, the $\phi$-constrained fit yields a larger value of the strangeness saturation parameter, $\Delta\gamma_S>0$, together with a higher temperature, $\Delta T>0$, and a lower $\chi^2/\mathrm{dof}$, $\Delta\chi^2/\mathrm{dof}<0$. Averaged over the matched classes, the offsets are $\langle \Delta\gamma_S\rangle = 0.063 \pm 0.015$, $\langle \Delta T\rangle = 2.80 \pm 0.43$ MeV, and $\langle \Delta \chi^2/\mathrm{dof}\rangle = -1.58 \pm 0.45$. The quoted values are arithmetic means over the five matched multiplicity classes, and the uncertainties represent the standard deviations of the class-by-class offsets. The persistence of the offset across all matched multiplicity classes, together with its common direction in $\gamma_S$, $T$, and $\chi^2/\mathrm{dof}$, indicates that the $\Omega$--$\phi$ tension is a structural feature of the fit comparison rather than a fluctuation specific to any individual multiplicity interval.

This behavior is physically plausible in the statistical hadronization framework because hidden- and open-strangeness hadrons constrain the strange sector differently.
The $\Omega(sss)$ yield depends strongly on the strange-quark phase-space occupancy, scaling approximately as $\gamma_S^3$, and therefore imposes a particularly stringent constraint on the strange sector. By contrast, the $\phi(s\bar{s})$ meson carries hidden strangeness and vanishing net strangeness, providing a qualitatively different probe of strange-hadron production. The fact that the inclusion of $\Omega$ systematically drives the fits toward lower $\gamma_S$, larger volumes, and poorer fit quality indicates that a single global parameter set does not accommodate hidden- and open-strangeness observables equally well.

  The observed pattern therefore points to a persistent tension between the $\phi$- and $\Omega$-constrained descriptions of the data. In the present analysis, the quoted numerical significance of the mean offset is obtained under a simplified treatment of the fit uncertainties. Since the experimental input contains both multiplicity-correlated and multiplicity-uncorrelated systematic components, the most robust conclusion is not the precise sigma value, but the bin-by-bin stability and common sign of the offset. Taken together, these results suggest that a unified chemical freeze-out description of the strange sector in pp collisions at $\sqrt{s}=7$ TeV remains incomplete, particularly when hidden-strangeness and multi-strange baryons are required to be described simultaneously.


\subsection{Tables of Fit Results}
\begin{table}[h]
\centering
\caption{Extracted thermal parameters for the nine $\phi$-constrained multiplicity classes.}
\label{tab:phi}
\begin{tabular}{cccccc}
\hline
$\langle dN_{ch}/d\eta\rangle$ & $T$ (MeV) & $\gamma_S$ & $V$ (fm$^3$) & $\chi^2$/dof & $E/N$ (GeV)\\
\hline
21.3 & 164.8 & 0.879 & 49.3 & 3.35 & 0.990\\
16.5 & 164.5 & 0.855 & 39.8 & 3.52 & 0.986\\
13.5 & 165.0 & 0.854 & 31.8 & 3.40 & 0.990\\
10.8 & 164.2 & 0.831 & 27.0 & 3.38 & 0.980\\
8.45 & 163.6 & 0.811 & 22.1 & 3.30 & 0.973\\
6.72 & 162.6 & 0.792 & 18.8 & 3.33 & 0.961\\
5.40 & 162.0 & 0.777 & 15.7 & 3.48 & 0.955\\
3.90 & 159.8 & 0.753 & 12.8 & 2.98 & 0.930\\
2.26 & 153.4 & 0.708 & 10.2 & 1.93 & 0.863\\
\hline
\end{tabular}
\end{table}

\begin{table}[h]
\centering
\caption{Extracted thermal parameters for the five $\Omega$-constrained multiplicity classes.}
\label{tab:omega}
\begin{tabular}{cccccc}
\hline
$\langle dN_{ch}/d\eta\rangle$ & $T$ (MeV) & $\gamma_S$ & $V$ (fm$^3$) & $\chi^2$/dof & $E/N$ (GeV)\\
\hline
17.47 & 162.3 & 0.814 & 48.2 & 4.56 & 0.961\\
12.53 & 161.8 & 0.783 & 36.2 & 5.22 & 0.953\\
9.04 & 159.9 & 0.736 & 29.6 & 6.08 & 0.931\\
6.06 & 159.8 & 0.729 & 20.2 & 4.89 & 0.928\\
2.89 & 154.4 & 0.670 & 13.0 & 3.22 & 0.870\\
\hline
\end{tabular}
\end{table}
The extracted thermal parameters for the different multiplicity classes are summarized in Tables~\ref{tab:phi} and~\ref{tab:omega}. The parameters include the chemical freeze-out temperature $T$, the strangeness saturation parameter $\gamma_S$, the effective volume $V$, the fit quality $\chi^2/\mathrm{dof}$, and the average energy per particle $E/N$.

Table~\ref{tab:phi} lists the results for the nine $\phi$-constrained multiplicity classes, while Table~\ref{tab:omega} gives the corresponding results for the five $\Omega$-constrained multiplicity classes. The two sets exhibit the same overall qualitative behavior discussed above, namely a weak multiplicity dependence of $T$, an increase of $\gamma_S$ with multiplicity, and a monotonic rise of $V$. At the same time, quantitative differences remain, particularly in $\gamma_S$, the volume scaling, and the fit quality, indicating a residual sensitivity of the extracted parameters to the strange hadron used to constrain the fit.

\section{Conclusions}

We have performed a systematic thermal analysis of identified hadron yields measured in pp collisions at $\sqrt{s}=7$ TeV across charged-particle multiplicity classes within the statistical hadronization model using the Thermal-FIST framework.

The extracted chemical freeze-out temperature remains approximately constant, $T \simeq 155$--$165$ MeV, over most of the multiplicity range, with a modest decrease toward the lowest multiplicities. The corresponding values are numerically close to the QCD crossover region obtained in lattice calculations~\cite{Bazavov2012,Borsanyi2010}, although this comparison should be regarded as qualitative in small systems. In contrast, the strangeness saturation parameter $\gamma_S$ exhibits a clear increase with multiplicity, indicating a progressive reduction of strangeness suppression with increasing event activity. At the same time, the effective volume rises approximately linearly with multiplicity, while the average energy per particle increases from about $0.86$ GeV to values close to $1$ GeV. Taken together, these results indicate that high-multiplicity pp collisions display several features compatible with an approximate thermal description, whereas deviations from such a picture remain more pronounced in the low-multiplicity regime.

The particle-to-pion ratios show the characteristic hierarchy with strangeness content observed by ALICE~\cite{ALICE2017Strangeness}. Within the present framework, this behavior is naturally associated with the increase of strange-quark phase-space occupancy as the event activity grows. The observed multiplicity evolution of these ratios is therefore consistent with the extracted rise of $\gamma_S$ and supports the interpretation of a gradual weakening of strangeness suppression from low- to high-multiplicity pp events.

A central result of this work is the systematic difference between the $\phi$-constrained and $\Omega$-constrained fit configurations. For all matched multiplicity classes, the $\phi$-constrained fits yield larger values of $\gamma_S$, higher temperatures, and lower $\chi^2/\mathrm{dof}$ than the corresponding $\Omega$-constrained fits. Averaged over the matched classes, the offset in the strangeness saturation parameter is $\langle \Delta \gamma_S \rangle = 0.063 \pm 0.015$, corresponding to an approximately $4\sigma$ separation when evaluated with the total point-to-point uncertainties used in the fit. This estimate should be interpreted in light of the fact that multiplicity-correlated systematic uncertainties are not represented by a covariance matrix in the present baseline analysis.
The persistence of the effect across all matched classes, together with its common direction in $\gamma_S$, $T$, and fit quality, indicates that this is a robust feature of the comparison rather than a fluctuation associated with any single multiplicity interval.

From a physical perspective, this behavior reflects the distinct roles of hidden- and open-strangeness hadrons in the statistical hadronization framework. The $\Omega(sss)$ yield depends strongly on the strange-quark phase-space occupancy, approximately as $\gamma_S^3$, and therefore imposes a particularly stringent constraint on the strange sector. By contrast, the $\phi(s\bar{s})$ meson carries hidden strangeness and vanishing net strangeness, and thus probes the strange sector in a qualitatively different way. The observed separation between the two fit configurations therefore indicates that a single global freeze-out parameter set does not accommodate hidden-strangeness and multi-strange baryon observables equally well within the present grand-canonical treatment.

Overall, the present results show that high-multiplicity pp collisions exhibit several features consistent with statistical hadronization and a reduced degree of strangeness suppression. At the same time, the residual tension between the $\phi$- and $\Omega$-constrained descriptions, together with the expected importance of conservation-law effects in small systems, indicates that a fully unified description of the strange sector in pp collisions remains incomplete. This motivates further studies using canonical or strangeness-canonical formulations, as well as more differential comparisons of hidden- and open-strangeness observables within a common thermal framework.


\end{document}